\documentclass {article} 

\usepackage [T2A] {fontenc} 
\usepackage [cp1251] {inputenc} 

\usepackage {amssymb, amsmath, amsthm, amsfonts} 
\usepackage {graphicx} 
\usepackage{hyperref}
\begin {document} 

\addcontentsline {toc} {section} {Adaptive two-dimensional wavelet transformation based on the Haar system} 
\begin {center} 
\bf M. Prisheltsev (Voronezh) \\
mikhail.prisheltsev@gmail.com \\
\uppercase {Adaptive two-dimensional wavelet transformation based on the Haar system} 
\end {center}

 \section {Introduction} 
 The purpose is to study qualitative and quantitative rates of image compression by using different Haar wavelet banks. The experimental results of adaptive compression are provided. The paper deals with specific examples of orthogonal Haar bases generated by multiresolution analysis. Bases consist of three piecewise constant wavelet functions with a support $[0,1] \times [0,1] $.
 
 \section {The theoretical part} 

 Let M be fixed $d \times d$ matrix of integers, such that absolute value of all of its eigenvalues is greater than 1. \

 \textbf {Definition 1 [1, p. 94].} A collection of closed spaces $ V_j \subset L_2 (\mathbb {R} ^ d), j \in \mathbb {Z} $ is called \textit {multiresolution analysis} (MRA) in the $ L_2 (\mathbb {R} ^ d) $ with matrix expansion coefficient $ M $, if the following conditions hold: 
1) $ V_j \subset V_ {j + 1} $ for all $ j \in \mathbb {Z} $; 
2) $ \bigcup_ {j \in \mathbb {Z}} V_j $ is dense in $ L_2 (\mathbb {R} ^ d) $; 
3) $ \bigcap_ {j \in \mathbb {Z}} V_j = \{0 \} $; 
4) $ f \in V_0 \Leftrightarrow f (M ^ j \cdot) \in V_j $ for all $ j \in \mathbb {Z} $; 
5) there is a function $ \varphi \in V_0 $ such that the sequence $ \{\varphi (\cdot + n) \} _ {n \in Z ^ d} $ forms a Riesz basis in $ V_0 $. 

 \textbf {Definition 2 [1, p. 141].} Let E be measurable set in $ \mathbb {R} ^ d $, such that $ \bigcup_ {l \in \mathbb {Z} ^ d} (E + l) = \mathbb {R} ^ d $, and the system $ \{\chi_E (\cdot + n) \} _ {n \in \textbf {Z} ^ d} $, where $ \chi_E $ is characteristic function of $ E $, is orthonormal. MRA generated by the function $ \chi_E $ is called \textit {Haar MRA}. \\

 Let $ \Psi $ be finite set of functions in $ L_2 (\mathbb {R} ^ d) $. A family of functions $ \eta (\Psi): = \{\psi_ {j, k}: = 2 ^ {jd / 2} \psi (2 ^ j \cdot-k), \psi \in \Psi, j \in \mathbb {Z}, k \in \mathbb {Z ^ d} \} $ we will call \textit {wavelet system} generated by the set of functions $ \Psi $. \\

\textbf {Theorem.} Let $ (V_j) _ {j \in \mathbb {Z}} \subset L_2 (\mathbb {R} ^ d) $ be Haar MRA with a scaling function 
$$ 
\chi _ {[0.1] ^ 2} (x, y): = 
\left \{\begin {matrix} 
 1, & \textit {if } (x, y) \in [0,1] \times [0,1], \\
 0, & \textit {if } (x, y) \notin [0,1] \times [0,1] 
\end {matrix} \right. $$ 

and with a matrix coefficient of expansion 
$ M = 
\begin {pmatrix} 
 2 & 0 \\
 0 & 2 
\end {pmatrix} 
$ 
and let $ \Theta $ be a piecewise constant wavelet system generated by the a set of wavelet functions 
$$ 
\eta_i (x, y) = 
\left \{\begin {matrix} 
a_ {i1}, & \textit {if } (x, y) \in [0,1 / 2] \times [0,1 / 2] \\
a_ {i2}, & \textit {if } (x, y) \in [0,1 / 2] \times [1 / 2,1] \\
a_ {i3}, & \textit {if } (x, y) \in [1 / 2,1] \times [1 / 2,1] \\
a_ {i4}, & \textit {if } (x, y) \in [1 / 2,1] \times [0,1 / 2] 
\end {matrix} \right. (1) 
$$ 

$ i = \overline {1,3} $, defined on a square $ [0,1] \times [0,1] $, where $ a_ {ij} \in \mathbb {R}$ are real numbers, $j = \overline {1 , 4}, i = \overline {1,3}$. Then \\

either following equalities are true for $ a_ {ij} $ 
 $$ a_ {11} = a_ {12} = a_ {14} = \lambda, a_ {13} = -3 \lambda, \textit {where} \lambda \in \mathbb {R} $$ 
 $$ a_ {i4} = 0, a_ {i3} = -a_ {i1} - a_ {i2}, i = 2,3 $$ 
 $$ 2a_ {21} a_ {31} + 2a_ {22} a_ {32} + a_ {21} a_ {32} + a_ {22} a_ {31} = 0 $$ 

i.e. four parameters are enough to calculate all the $ a_ {ij}, j = \overline {1,4}, i = \overline {1,3} $, \\

or 
$$ a_{i1} = \frac {cos \alpha_i} {\sqrt {3}} - \frac {2sin \alpha_i cos \beta_i} {\sqrt {6}}, $$ 
$$ a_{i2} = \frac {cos \alpha_i} {\sqrt {3}} + \frac {sin \alpha_i cos \beta_i} {\sqrt {2}} + \frac {sin \alpha_i sin \beta_i} {\sqrt {6}}, \\$$ 
$$ a_{i3} = \frac {cos \alpha_i} {\sqrt {3}} - \frac {sin \alpha_i cos \beta_i} {\sqrt {2}} + \frac {sin \alpha_i sin \beta_i} {\sqrt {6}}, $$

where $ 0 \leq \alpha_i \leq \pi $ and $ 0 \leq \beta_i \leq 2 \pi$ are the parameters with independent $ \alpha_2, \alpha_3 $ only, $i = \overline {1,3} $. 

\textbf {Corollary.} The set of wavelet functions (1) with real values where only one coeffceint for each function is equal to zero can have only one of the following forms: 
 1) $ a_ {14} = a_ {21} = a_ {32} = 0 $, 2) $ a_ {12} = a_ {23} = a_ {34} = 0 $, 3) $ a_ {11} = a_ {22} = a_ {33} = 0 $ 4) $ a_ {13} = a_ {24} = a_ {31} = 0 $. \\

\textbf {Note.} The book [1] considers the following approach to the construction of wavelet functions. We need some defenitions to describe it. 

1. Let A be a $ d \times d $ non-singular matrix of integers,than the vectors $ k, n \in \textbf{Z}^d $ are congruent modulo A, if $ k - n = Al, l \in \textbf{Z}^d $. Integer lattice $\textbf{Z}^d$ is partitioned into cosets regarding this congruent relation. The set containing exactly one representative of each coset is called the set of numbers of the matrix A [1, p. 89]. 

2. Let $ \varphi $ be scaling function for a MRA. Then the refinement equation holds 

$$ \varphi = \sum_ {n \in \textbf {Z} ^ d} h_nm ^ {1/2} \varphi (M \cdot + n), \sum_ {n \in \textbf {Z} ^ d} | h_n ^ 2 | <\infty $$ 

or in the Fourier notaion
$$ \hat {\varphi} (\xi) = m_0 (M ^ {* - 1} \xi) \hat {\varphi} (M ^ {* - 1} \xi) $$
, where $ m_0 (\xi) = m ^ {- 1/2} \sum_ {n \in \textbf {Z} ^ d} h_n e ^ {2 \pi i (n, \xi)} $ is the mask [1, c .94]. 

\textbf {Statement} [1. c.94]. Let MRA $ (V_j) _ {j \in \mathbb {Z}} $ be generated by scaling function $ \varphi $ with a mask $ m_0 $ and the system $ \{\varphi (\cdot + n) \} _ {n \in \mathbb {Z} ^ d} $ be orthonormal. If $ \psi ^ {(\nu)} (\nu = 1, ..., m-1)$ is a set of wavelet functions defined by the equations $ \hat {\psi} ^ {(\nu)} (\xi ) = m_ \nu (M ^ {* - 1} \xi) \hat {\varphi} (M ^ {* - 1} \xi) $ with the masks $ m_ \nu, \nu = 1, .. ., m-1 $, such that for almost all $ \xi \in \mathbb {R} ^ d $ the matrix 
$$ M: = \{m_ \nu (\xi + M ^ {* - 1} s_k) \} _ {\nu, k = 0} ^ {m-1} $$ 

is unitary ($ \{s_0, ..., s_ {m-1} \} $ is an arbitrary set of numbers of the matrix $ M ^ {*} $), then the functions $ \psi ^ {(\nu)}, \nu = 1 , ..., m-1 $ form a set of wavelet functions. 

If we apply this approach to the above-described case of the Haar MRA with scaling function $ \chi _ {[0,1] ^ 2} (x, y) $ and the matrix coefficient of expansion 
$ M = \begin {pmatrix} 
2 & 0 \\
0 & 2 
\end {pmatrix} $ 
we'll get that the function (1) must satisfy the equations 
$$ \hat {\eta} _i (\xi) = m_i (M ^ {- 1} \xi) \hat {\chi} _ {[0,1] ^ 2} (M ^ {- 1} \xi ), i = 1,2,3, \xi = (\xi_1, \xi_2) \in \mathbb {R} ^ 2.$$

 \section {Haar bases in $ L_2 (\textbf {R} ^ d) $} 

 \textbf {Definition 3.} A set of wavelets $ \{\psi ^ 1, \psi ^ 2, \psi ^ 3 \} $ given in tabular form (set1) is called a \textit {(classical) Haar basis}. This set is a orthonormal basis (ONB) in $ L ^ 2 (\textbf {R} ^ 2) $.

\begin{table}[h]
\centering
\begin{tabular}{ccccccccll}
\cline{1-2} \cline{4-5} \cline{7-8}
\multicolumn{1}{|c|}{1} & \multicolumn{1}{c|}{-1} & \multicolumn{1}{c|}{} & \multicolumn{1}{c|}{1} & \multicolumn{1}{c|}{1} & \multicolumn{1}{c|}{} & \multicolumn{1}{c|}{1} & \multicolumn{1}{c|}{-1} &  &  \\ \cline{1-2} \cline{4-5} \cline{7-8}
\multicolumn{1}{|c|}{1} & \multicolumn{1}{c|}{-1} & \multicolumn{1}{c|}{} & \multicolumn{1}{c|}{-1} & \multicolumn{1}{c|}{-1} & \multicolumn{1}{c|}{} & \multicolumn{1}{c|}{-1} & \multicolumn{1}{c|}{1} &  & (set1) \\ \cline{1-2} \cline{4-5} \cline{7-8}
\\
\multicolumn{2}{c}{$\psi^1$} &  & \multicolumn{2}{c}{$\psi^2$} &  & \multicolumn{2}{c}{$\psi^3$} & &
\end{tabular}
\end{table}

 Now we construct 3 alternative (ONB): 
 \begin {itemize} 
 \item "Vertical" (set2) 
 \item "Horizontal" (set3) 
 \item "Diagonal" (set4) 
 \end {itemize}

\begin{table}[h]
\centering
\begin{tabular}{ccccccccll}
\cline{1-2} \cline{4-5} \cline{7-8}
\multicolumn{1}{|c|}{1} & \multicolumn{1}{c|}{-1} & \multicolumn{1}{c|}{} & \multicolumn{1}{c|}{1} & \multicolumn{1}{c|}{0} & \multicolumn{1}{c|}{} & \multicolumn{1}{c|}{0} & \multicolumn{1}{c|}{1} &  &  \\ \cline{1-2} \cline{4-5} \cline{7-8}
\multicolumn{1}{|c|}{1} & \multicolumn{1}{c|}{-1} & \multicolumn{1}{c|}{} & \multicolumn{1}{c|}{-1} & \multicolumn{1}{c|}{0} & \multicolumn{1}{c|}{} & \multicolumn{1}{c|}{0} & \multicolumn{1}{c|}{1} &  & (set2) \\ \cline{1-2} \cline{4-5} \cline{7-8}
\\
\multicolumn{2}{c}{$\psi_v^1$} &  & \multicolumn{2}{c}{$\psi_v^2$} &  & \multicolumn{2}{c}{$\psi_v^3$} & &
\end{tabular}
\end{table}

\begin{table}[h]
\centering
\begin{tabular}{ccccccccll}
\cline{1-2} \cline{4-5} \cline{7-8}
\multicolumn{1}{|c|}{1} & \multicolumn{1}{c|}{-1} & \multicolumn{1}{c|}{} & \multicolumn{1}{c|}{1} & \multicolumn{1}{c|}{-1} & \multicolumn{1}{c|}{} & \multicolumn{1}{c|}{0} & \multicolumn{1}{c|}{0} &  &  \\ \cline{1-2} \cline{4-5} \cline{7-8}
\multicolumn{1}{|c|}{1} & \multicolumn{1}{c|}{-1} & \multicolumn{1}{c|}{} & \multicolumn{1}{c|}{0} & \multicolumn{1}{c|}{0} & \multicolumn{1}{c|}{} & \multicolumn{1}{c|}{1} & \multicolumn{1}{c|}{-1} &  & (set3) \\ \cline{1-2} \cline{4-5} \cline{7-8}
\\
\multicolumn{2}{c}{$\psi_h^1$} &  & \multicolumn{2}{c}{$\psi_h^2$} &  & \multicolumn{2}{c}{$\psi_h^3$} & &
\end{tabular}
\end{table}

\begin{table}[h]
\centering
\begin{tabular}{ccccccccll}
\cline{1-2} \cline{4-5} \cline{7-8}
\multicolumn{1}{|c|}{1} & \multicolumn{1}{c|}{-1} & \multicolumn{1}{c|}{} & \multicolumn{1}{c|}{1} & \multicolumn{1}{c|}{0} & \multicolumn{1}{c|}{} & \multicolumn{1}{c|}{0} & \multicolumn{1}{c|}{1} &  &  \\ \cline{1-2} \cline{4-5} \cline{7-8}
\multicolumn{1}{|c|}{1} & \multicolumn{1}{c|}{-1} & \multicolumn{1}{c|}{} & \multicolumn{1}{c|}{0} & \multicolumn{1}{c|}{-1} & \multicolumn{1}{c|}{} & \multicolumn{1}{c|}{-1} & \multicolumn{1}{c|}{-1} &  & (set4) \\ \cline{1-2} \cline{4-5} \cline{7-8}
\\
\multicolumn{2}{c}{$\psi_d^1$} &  & \multicolumn{2}{c}{$\psi_d^2$} &  & \multicolumn{2}{c}{$\psi_d^3$} & &
\end{tabular}
\end{table}

 \section {Adaptive Haar transform} 

 \textbf {Definition 4.} The 4 numbers $a, v, h, d$ such that 
 $$ a = (m_ {11} + m_ {12} + m_ {21} + m_ {22}) / 4, $$ 
 $$ v = \sum_ {i = 1} ^ {2} \sum_ {j = 1} ^ {2} \psi_ {ij} ^ 1 m_ {ij}, $$ 
 $$ h = \sum_ {i = 1} ^ {2} \sum_ {j = 1} ^ {2} \psi_ {ij} ^ 2 m_ {ij}, $$ 
 $$ d = \sum_ {i = 1} ^ {2} \sum_ {j = 1} ^ {2} \psi_ {ij} ^ 3 m_ {ij}. $$ 
  are called Haar transform of a matrix M (dim M = 2).

 This wavelet transform we will denote by $ \{\psi ^ 1, \psi ^ 2, \psi ^ 3 \}: M \Longrightarrow (a, v, h, d) $.

\textbf {Definition 5.} \textit {Adaptive Haar transform} is four numbers, such that 

\begin {align} 
(a ^ *, v ^ *, h ^ *, d ^ *) = \{(a, v_k, h_k, d_k) | \{\psi ^ 1, \psi ^ 2, \psi ^ 3 \} _ {set k}: M \Longrightarrow \nonumber \\
 (a, v_k, h_k, d_k), (v_k ^ 2 + h_k ^ 2 + d_k ^ 2) \underset {k = \overline {1,4}} {\rightarrow} min \}. \nonumber 
\end {align} 

In other words we perform 4 Haar transformation of matrix M and choose the one, which minimizes the sum of squares of coefficients v, h, d.

 \section {The adaptive image compression} 

Let M be $n \times m$ matrix. Performing a two-dimensional wavelet transformation by a Haar basis, we'll get 4 $ n / 2 \times m / 2 $ matricies A, V, H, D. 

\textbf {Definition 6.} The wavelet transformation
$ \{A, V, H, D | \left \| V \right \| + \left \| H \right \| + \left \| D \right \| \rightarrow min \} $, where $ \left \| M \right \| = \sum_ {j = 1} ^ {} m_ {ij} ^ 2 $ we'll call an adaptive Haar transformation of matrix M.

To construct the compression algorithm, we'll use JPEG2000 as a template, replacing wavelet transformation by our adaptive Haar transformation. The other difference is a maintenance of the id of the used basis (set 1 - 4). Schematically, the image compression and restoration process is presented in the \textit {Figure 1}. 

\begin {figure} [h] 
\centering 
\includegraphics [width = \textwidth] {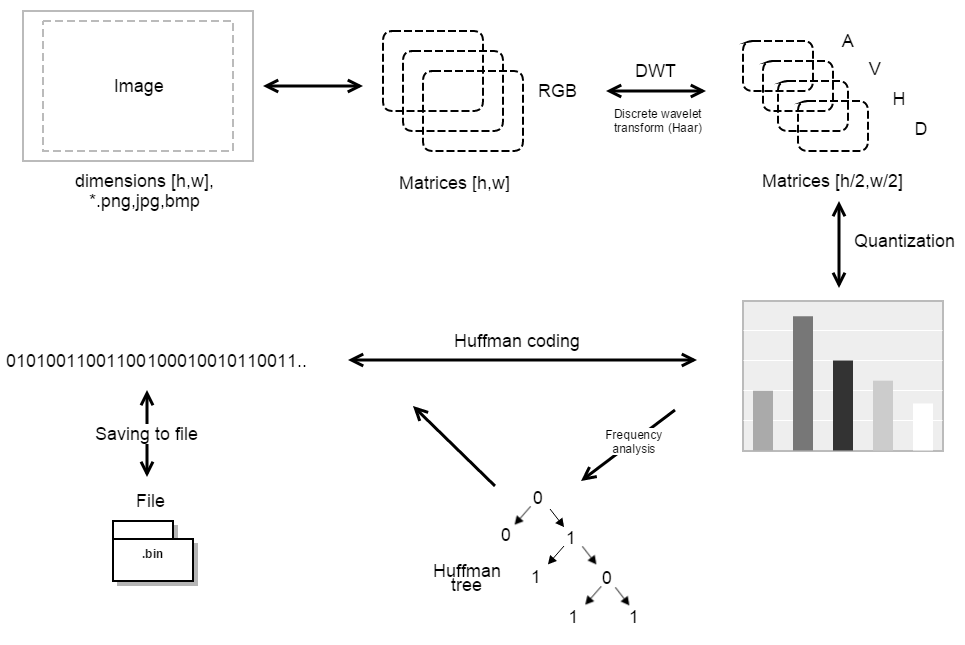} 
 Scheme 1.
\end {figure}

 Steps: 
\begin {enumerate} 
 \item 
Converting the image to three matrices (RGB colors) with elements רע [0, 255], the matrices dimension coincides with the dimension of the image. 
\item Wavelet transformation of each of the color matrices. As a result we'll get 4 matrices twice smaller than the original (with elements in [-255, 255]). 
\item Quantization. At this step the major compression takes place, and hence there is a loss of information (picture quality). 
\item Huffman processing. Frequency analysis and compression, conversion to binary code. 
\item Assembling binary codes into a single structure, i.e. binary file. 

\end {enumerate}

 \section {Results} 

Analysing experimental results of the implemented software application, we focused on the following input parameters: 

\begin {itemize} 
\item image (type, size, color, etc.) 
\item quantization levels quantity(fewer levels mean the better compression) 
\item wavelet decomposition level (more levels mean the better compression efficiency of quantization, also more errors in the recovery, the worse quality) 
\item wavelet type (adaptive or specific Haar basis) 
\end {itemize} 

The final compression rates for different parameters are in the range of 30 \% - 50 \%. \\
Quantization: 64 levels - 54 \%, 32l - 49 \%, 16l - 45 \%, 8l - 44 \%. \\
The level of decomposition: 1 - 49 \%, 2 - 37 \% 3 - 35 \%, 4 - 34 \%. \\
The wavelet type: an adaptive scheme is worse than simple wavelet bases compression in average by 1-3 \% due to the fact that the basis id matrix is also stored (id of used wavelet). \\
The correlation between the type of image and compression was not revealed.

The processed images are available via the link \url{http://goo.gl/J0hC3B}. The source code of the application you can find on \url{https://github.com/mihin/haar-image-compression}.

Further analysis of the developed filter bank will be based on the paper [4]. \\

I wish to thank Igor Novikov for posing the problem and the attention to the research. \\

This research was carried out with the financial support of the Russian Foundation for Basic Research (grant no.~14-31-50222).

\section {References}

1. \textit {Novikov I.Ya., Protasov V.Yu., Skopina M.A.} Wavelet theory. M .: Fizmatlit, 2005. 616 c. 

2. \textit {Hur Y., Ron A.} New constructions of piecewiseconstant wavelets // Electronic Transactions on Numerical Analysys. 2006. Vol. 25. P. 138-157. 

3 \textit {Krasilnikova M.S.} parameterization of the two-dimensional non-separable Haar wavelets. // Sarat. Univ. 2011. 11 Ser. Mathematics. Mechanics. Computer Science, vol. 2. p. 26-32. 

4 \textit {M.K. Tchobanou} Multirate digital signal processing system. M .: Publishing House MEI, 2009. 120 c. 

5. \textit {S. Mallat.} A Wavelet Tour of Signal Processing. Academic Press, 1998. 

6. \textit {M. Unser.} Mathematical Properties of the JPEG2000 Wavelet Filters. // IEEE Transactions on Image Processing, v. 12, 9, 2003. P.1080--1090.

\end {document}